\documentclass[12pt]{spieman}  % 12pt font required by SPIE;
\usepackage{amsmath,amsfonts,amssymb}
\usepackage{graphicx}
\usepackage{setspace}
\usepackage{tocloft}
\usepackage{xcolor}
\title{Fabrication and characterization of optical nanofiber tips}

\author[a]{Resmi M}
\author[a]{Elaganuru Bashaiah}
\author[a,*]{Ramachandrarao Yalla}
\affil[a]{School of Physics, University of Hyderabad, Hyderabad, Telangana, India, 500046}
\cftpagenumbersoff{figure}
\cftpagenumbersoff{table} 
\begin{document} 
\maketitle

\begin{abstract}
We demonstrate the fabrication of an optical nanofiber tip (ONFT) using a two-step chemical etching technique. This technique employs 30\% and 24\% hydrofluoric (HF) acid for the first and second steps, respectively. In the first step, a silica single-mode fiber with clad diameter of 125 $\mu m$ and core diameter of 10 $\mu m$ is immersed in the HF acid for 90 minutes. The resultant fiber diameter is reduced to $\sim$45 $\mu m$. In the second step, such a fiber is again immersed in the HF acid for 60-80 minutes. The etching time is controlled to achieve the desired tip diameter of the ONFT. We characterize fabricated ONFTs by measuring optical transmission and surface morphology. The observed optical transmissions are more than 30\% and tip diameters are less than 500 $nm$. The measured results readily reveal the merit of the employed technique as it is easy and cost-effective. Due to the strong confinement of the electromagnetic field at the tip of ONFTs, these structures lay versatile platforms with potential applications in sensing, photonics, and quantum optics.
\end{abstract}

% Include a list of up to six keywords after the abstract

% Include email contact information for corresponding author
{\noindent \footnotesize\textbf{*}Ramachandrarao Yalla,  \linkable{rrysp@uohyd.ac.in} }

\begin{spacing}{1}   % use double spacing for rest of manuscript

\section{Introduction}
\label{sect:intro}  % \label{} allows reference to this section
A tapered optical fiber with a sub-wavelength-diameter tip is defined as an optical nanofiber tip (ONFT). Due to high contrast between effective refractive indices of its core (silica) and clad (air), the ONFT offers strong field confinement \cite{tong2004single,tong2011subwavelength}, leading to wide range of applications in fields of photonics \cite{xiong2020multifunctional,wu2013optical}, nanophotonics \cite{tiecke2015efficient,tong2003subwavelength}, quantum optics \cite{decombe2013single,li2018fibre,brambilla2010optical,resmi2023efficient}, and sensing \cite{tong2018micro,berneschi2020optical,paiva2018optical,lu2000nanoscale}. Nowadays, the ONFT has become a promising candidate due to its utilization in various research fields. However, the fabrication of a high optical-quality ONFT is still a challenging task. For the fabrication of the ONFT and optical nanofibers, various techniques have been proposed and demonstrated \cite{lee2019fabrication,ward2006heat,mukhtar2010fabrication,lazarev2003formation,grosjean2007fiber,kbashi2012fabrication}. Examples would include flame-brushing and pulling \cite{lee2019fabrication}, CO$_{2}$ laser-assisted heating and pulling \cite{ward2006heat}, electric arc heating and one-sided pulling \cite{mukhtar2010fabrication}, laser-heated pulling and bending \cite{lazarev2003formation}, mechanical polishing \cite{grosjean2007fiber}, and chemical etching \cite{kbashi2012fabrication}. Additionally, a combination of electric arc heated micro-pulling with chemical etching \cite{huo2006fabrication} and melt-stretching with chemical etching \cite{ren2007preparation} have been demonstrated experimentally.

In the flame brushing and pulling method \cite{lee2019fabrication}, Lee \textit{et al.} have demonstrated the fabrication of a tapered fiber length of 217 $mm$, diameter of 820 $nm$, and transmission of $\sim$90\% (at a wavelength of 1550 $nm$). In the CO$_{2}$ laser-assisted heating and pulling \cite{ward2006heat}, Ward \textit{et al.} have demonstrated the fabrication of a fiber tapered diameter of $3$ - $4$ $\mu m$ with the transmission of $\sim$93\% (at a wavelength of 980 $nm$). In the electric arc heating and one-sided pulling technique \cite{mukhtar2010fabrication}, the fabrication of a fiber tip with a diameter of 1.4 $\mu m$ was demonstrated. In the laser-heated pulling and bending technique \cite{lazarev2003formation}, the fabrication of a fiber tip with a short tapering length of $\sim$300 $\mu m$ and a diameter of $\sim$50 $nm$ was demonstrated. In the chemical etching technique using HF acid \cite{kbashi2012fabrication}, the fabrication of the ONFT with a diameter of 592 $nm$ was demonstrated. 

Furthermore, a few modifications have been demonstrated in the chemical etching technique. The ONFT has been fabricated using HF acid and protective silicone oil on it \cite{nikbakht2015fabrication,puygranier2000chemical}. A syringe pump regulates the meniscus height on the etchant's surface, resulting in the desired cone angle. Dynamic selective etching was used to demonstrate a parabolic shape optical nanofiber probe and such a fiber tip shows improved transmission \cite{zhu2013dynamic}. Translating the fiber vertically during the etching process can control the shape of the fiber optic tips \cite{haber2004shape}. Researchers have used dynamic etching to fabricate a variety of tip shapes \cite{muramatsu1999dynamic}. Short and long tapered ONFTs have been fabricated by translating the fiber vertically upwards and downwards, respectively. 

Among the above-mentioned techniques, chemical etching is economical as no sophisticated instruments are required. It is easily adaptable and short tapered ONFTs can be experimentally realized. Additionally, mechanical stability of the ONFT is required for its application in different research fields for robust usage. For example, chemically etched fiber tips have been widely used for sensing \cite{decombe2013single,ascorbe2016high,zaca2018etched,chyad2015fabrication,cullum2000development,pathak2017fabrication,mondal2012optical} and quantum optics applications \cite{resmichanneling,shashank}. 
The ONFT with long tapering is mechanically unstable compared to a short tapered ONFT. Hence, the chemical etching technique for fabricating the short tapered ONFT with a sub-wavelength diameter tip draws attention. Although various groups worldwide have demonstrated the ONFT fabrication, the systematic fabrication and characterization of the ONFT have not been reported yet. 

In this paper, we demonstrate the fabrication of the ONFT using a two-step chemical etching technique. This technique employs 30\% and 24\% HF acid for the first and second steps, respectively. In the first step, a silica single-mode fiber (SMF) with 125 $\mu m$ clad and 10 $\mu m$ core diameters is immersed in the HF acid for 90 minutes. The resultant fiber diameter is reduced to $\sim$45 $\mu m$. In the second step, such a fiber is again immersed in the HF acid for 60 - 80 minutes. The measured optical transmissions are from 36\% to 95\%. The measured tip diameters of ONFTs are from 22 $nm$ to 820 $nm$.

\section{Fabrication procedure}
\label{sec:Exp}
\subsection{Pre-preparation process}
\label{sec:Pre}
Initially, we start with a 30 $cm$ long SMF (1550 BHP, Coherent) of single mode cut-off wavelength at 1.3 $\mu m$. This fiber consists of a core, clad, and outer plastic jacket with diameters 10 $\mu m$, 125 $\mu m$, and 250 $\mu m$, respectively. Using a three-hole fiber stripping tool (FTS4, Thorlabs), the outer jacket is carefully indented at one end, around 2 $cm$ from the edge. Subsequently, the indented end is immersed in acetone for $5$ - $10$ minutes to remove the outer plastic jacket without scratches on the fiber surface. The resultant fiber diameter is reduced to 125 $\mu m$. The fiber is meticulously cleaned with methanol to ensure it remains contaminant-free. The stripped end of the fiber is cleaved perpendicularly using a fiber cleaver (XL411, Thorlabs). The fiber is securely attached to a metal holder using a UV-curable glue. The metal holder consists of a V-groove on one end so the excess glue does not reach the cleaved fiber end. A small drop of the UV curable glue is deposited at the V-groove location. Subsequently, the UV light is directed to fix the fiber on the metal holder. Additionally, a scotch tape is fixed on the UV glue location for more stability. The metal holder with the SMF affixed on it, is placed on an \textit{xyz}-translation stage (TS 65, Holmarc) to achieve precise sample movement. Note that the fiber end is positioned downwards for immersing into the HF meniscus.
\subsection{Methodology for etching process}

\begin{figure}[!h]
	\centering
	\includegraphics[width=0.6\linewidth]{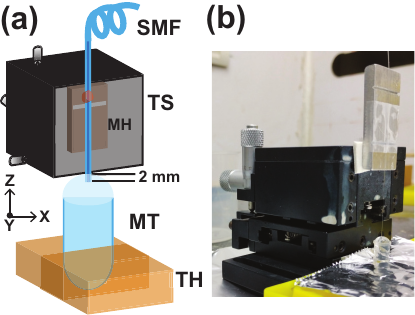}
	\caption{(a) A conceptual diagram for the fabrication of the optical nanofiber tip. SMF, TS, MH, MT, and TH denote single-mode fiber, translation stage, metal holder, microcentrifuge tube and tube holder, respectively. (b) The picture of the experimental scenario.}
	\label{fig1}
\end{figure}

A schematic of the experiment and the corresponding picture are shown in Figs. \ref{fig1} (a) and (b), respectively. The whole experiment is set up in a fume cupboard. A microcentrifuge tube of capacity 500 $\mu l$ is firmly fixed on a tube holder. The original 40\% HF acid (Avra, ASH2565) is diluted to 30\% and 24\% using de-ionized water. 30 $ml$ 40\% HF and 10 $ml$ de-ionized water is mixed to get 40 $ml$ of 30\% HF acid. Also 24 $ml$ 40\% HF and 16 $ml$ de-ionized water is mixed to get 40 $ml$ of 24\% HF acid. The translation stage is placed near the microcentrifuge tube such that the SMF end is just above the microcentrifuge tube. The microcentrifuge tube is filled with 30\% and 24\% HF acid in the first and second steps, respectively. Then, HF acid forms a meniscus on top of the microcentrifuge tube as clearly seen in Fig. \ref{fig1} (b). The length of the SMF dipped in the acid is precisely controlled by the translation stage. In the first step, 2 $mm$ of the fiber end is immersed in the acid for 90 minutes for the temperature ranging from 25-28$^{\circ}$C. The fiber is carefully withdrawn from the acid at the end. Again, the fiber is immersed in 24\% HF acid for 60/70/80 minutes. At the end, the fiber is withdrawn from the acid, and any residual acid is rinsed off with de-ionized water to avoid any further unintended etching of the fiber. The fabricated ONFT is preserved in a clean environment to avoid any contamination that can result in degradation. 
\section{Characterization of the fabricated ONFT samples}
We characterize optical and morphological properties of the fabricated ONFT using laser and field emission scanning electron microscopy (FESEM) (Sigma 360 VP, Zeiss).
\subsection{Optical}
\begin{figure}[!h]
	\centering
	\includegraphics[width=\linewidth]{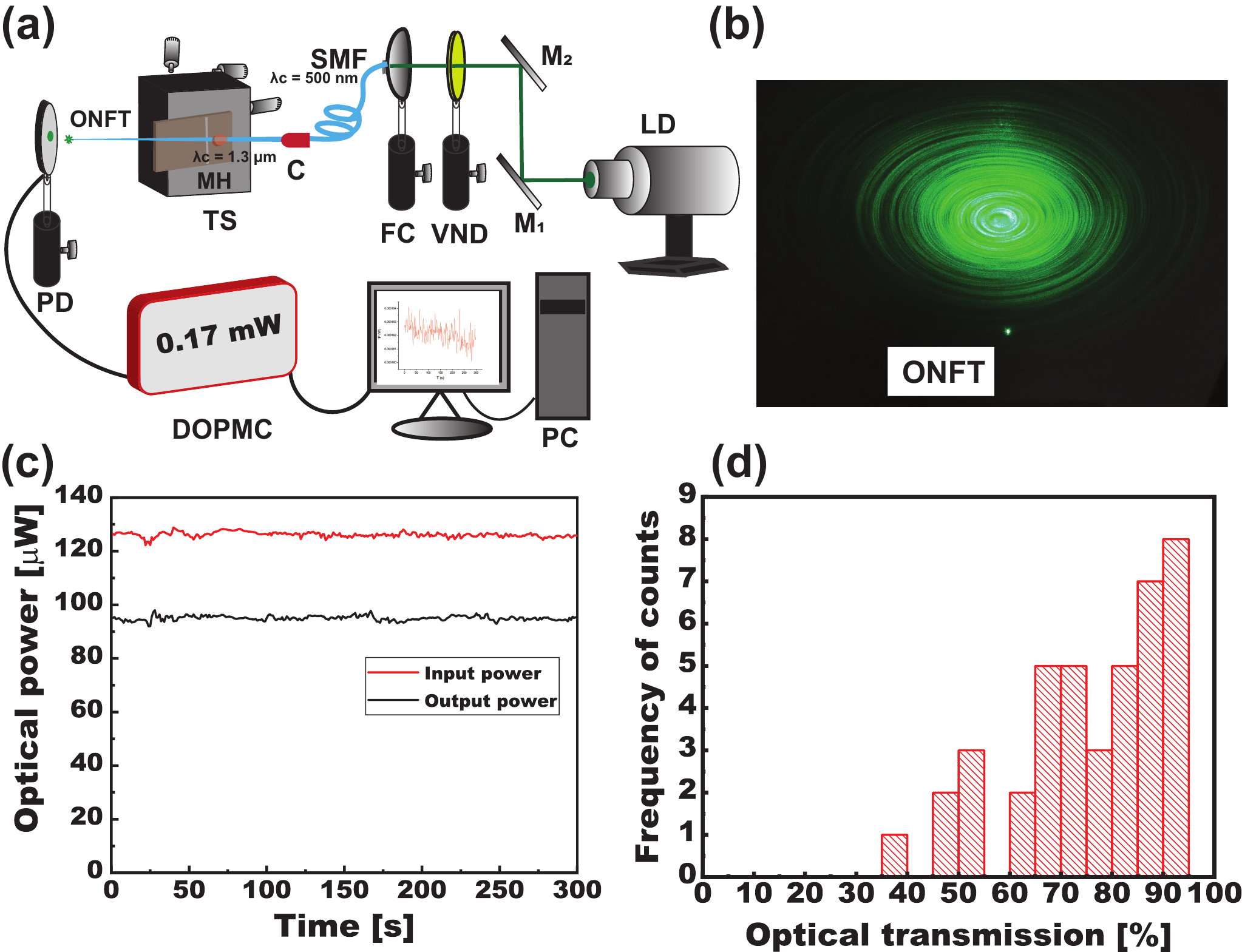}
	\caption{(a) A schematic of the experimental setup for optical transmission measurements. LD, M, VND, FC, SMF, TS, MH, PD, C, ONFT, and DOPMC denote laser diode, mirror, variable neutral density filter, fiber coupler, single mode fiber, translation stage, metal holder, photodiode, connector, optical nanofiber tip, and digital optical power meter console, respectively. (b) A typical diffraction pattern as soon as the laser enters into guided modes of the ONFT. (c) A typically measured input and output powers as a function of time. (d) A histogram profile for the measured optical transmissions through guided modes of ONFTs.}
	\label{fig2}
\end{figure}
The experimental sketch is shown in Fig. \ref{fig2} (a). The optical characterization of the ONFT is performed by monitoring the power transmitted through its guided modes. A laser diode (DL-G-5, Holmarc) at a wavelength of 532 $nm$ is used. The laser light is directed through two mirrors to a variable neutral density filter (HO-VND-N50, Holmarc) to vary the intensity of the laser light. A SMF patch cable and a fiber coupler guide the laser into the ONFT. The SMF patch cable has a single mode cut-off wavelength at 500-600 $nm$. Therefore it filters all other modes and allows only the fundamental mode to the ONFT. Single mode propogation is confirmed by observing the beam profile by connecting the SMF patch cable to the laser, by connecting bare fiber to the laser through the SMF patch cable before fabrication, and by connecting the ONFT to the laser after fabrication. As shown in Fig. \ref{fig2} (b), one can readily observe a diffraction pattern. The optical power measurements are carried out using a photodiode (S120VC, Thorlabs) connected to a digital optical power meter console (PM100D, Thorlabs). The output power is measured for 15 minutes by keeping the photodiode close to the tip of the ONFT. To measure the input power, the ONFT is cleaved perpendicularly at a safe distance from the tip, and the optical power at the cleaved end of the SMF is measured. Note that the input power is measured as close to the ONFT to avoid coupling loss from the SMF patch cable to the ONFT through the fiber coupler. The optical transmission is derived as the ratio of the output power to the input power. Figure \ref{fig2} (c) shows a typically measured input and output powers as a function of time. The black and red traces correspond to the output and input powers, respectively. The measured transmission value is 75.0 ($\pm$1.5)\%. We performed such transmission measurements for several samples. A summary of such measurements is plotted as a histogram and shown in Fig. \ref{fig2} (d). The bin width of the histogram is 5\%, which is greater than the error in the measurement. The histogram reflects that most ONFT samples exhibit optical transmission $\ge$ 60\%. The maximum/minimum measured optical transmission is $\sim$95\%/36\%. The perpendicularly cleaved ONFT, after the output power measurement, is used for the morphological characterization.

\subsection{Morphological}
\begin{figure}[!h]
	\centering
	\includegraphics[width=0.95\linewidth]{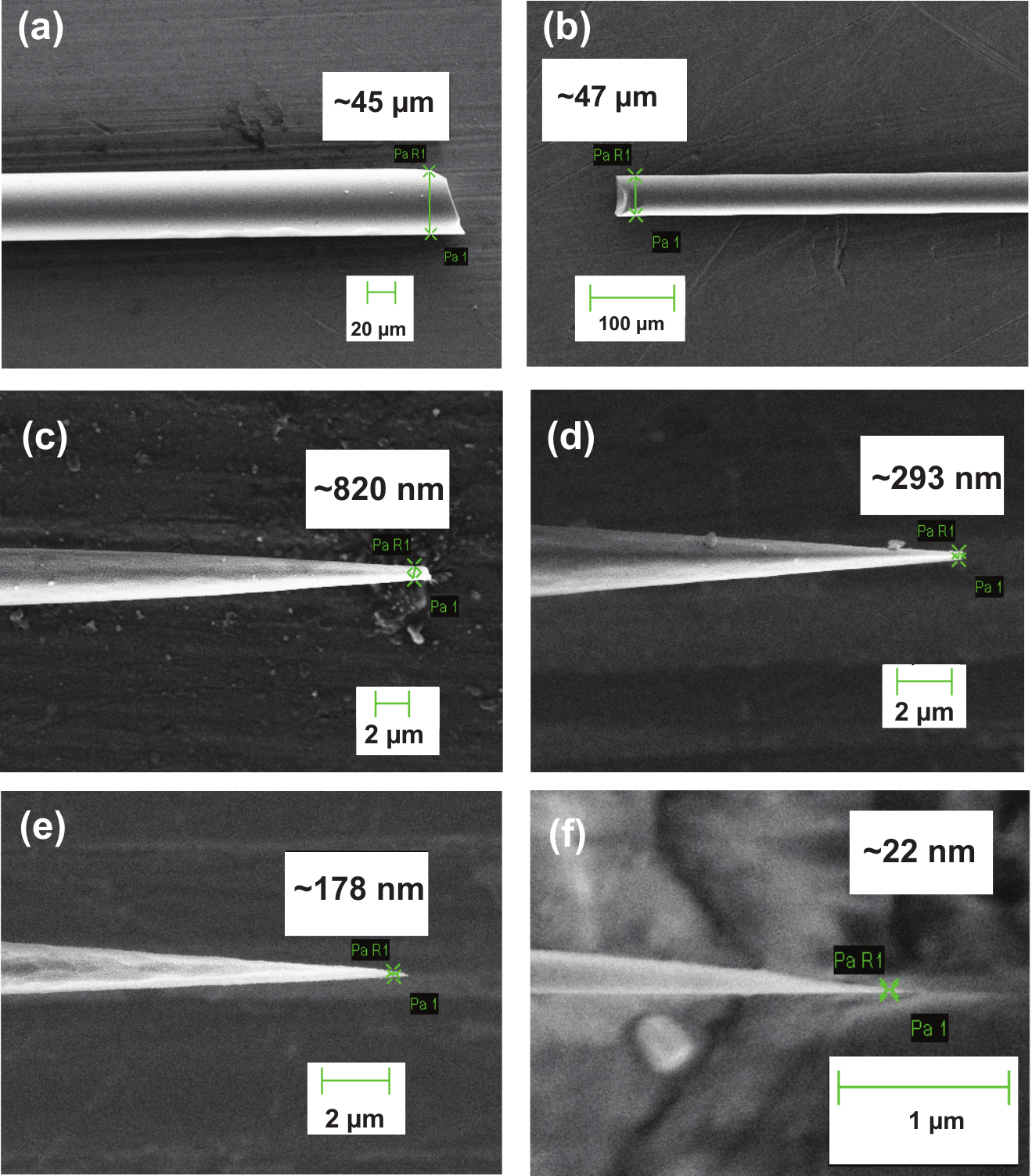}
	\caption{Typical field emission scanning electron microscopy images of a few optical fiber tips with different diameters. (a)-(b) and (c)-(f) are results after the first and second steps, respectively. The measured diameters of (a), (b), (c), (d), (e), and (f) are 45 ($\pm$1) $\mu m$, 47 ($\pm$2) $\mu m$, 820 ($\pm$50) $nm$, 293 ($\pm$70) $nm$, 178 ($\pm$50) $nm$, and 22 ($\pm$7) $nm$, respectively.}
	\label{fig3}
\end{figure}

The ONFT is affixed to the metal holder using the UV curable glue as explained in the section \ref{sec:Pre}. Note that the ONFT is not free-hanging but rests on the metal holder. To avoid charge-up, a mixture of gold and palladium is sputter coated on samples for 3 minutes, resulting 2-3 $nm$ layer on the surface of ONFTs. An accelerating voltage of 5 kV is applied for imaging. The tip of the ONFT is magnified, and the diameter is measured. After the first step, typical FESEM images of two samples are shown in Figs. \ref{fig3} (a) and (b). The measured diameters are 45 ($\pm$1) $\mu m$ and 47 ($\pm$2) $\mu m$ for (a) and (b), respectively.  After the second step, typical FESEM images of four ONFT samples are shown in Figs. \ref{fig3} (c)-(f). For (c), the measured ONFT diameter is 820 ($\pm$50)  $nm$ with etching timings of 90 and 70 minutes. The optical transmission of 47.0 ($\pm$1.5)\% is observed for it. For (d), the measured ONFT diameter is 293 ($\pm$70) $nm$ with etching timings 60 and 30 minutes. The observed optical transmission is 60 ($\pm$2)\%. For (e), the measured ONFT diameter is 178 ($\pm$50) $nm$ with etching timings 60 and 40 minutes. The measured optical transmission is 81 ($\pm$1)\%. For (f), the measured ONFT diameter is 22 ($\pm$7) $nm$ with etching timings 90 and 80 minutes. The optical transmission of 91 ($\pm$3)\% is observed for it. 
\pagebreak
\subsection{Data Analysis}
\begin{figure}[!h]
	\centering
	\includegraphics[width=0.9\linewidth]{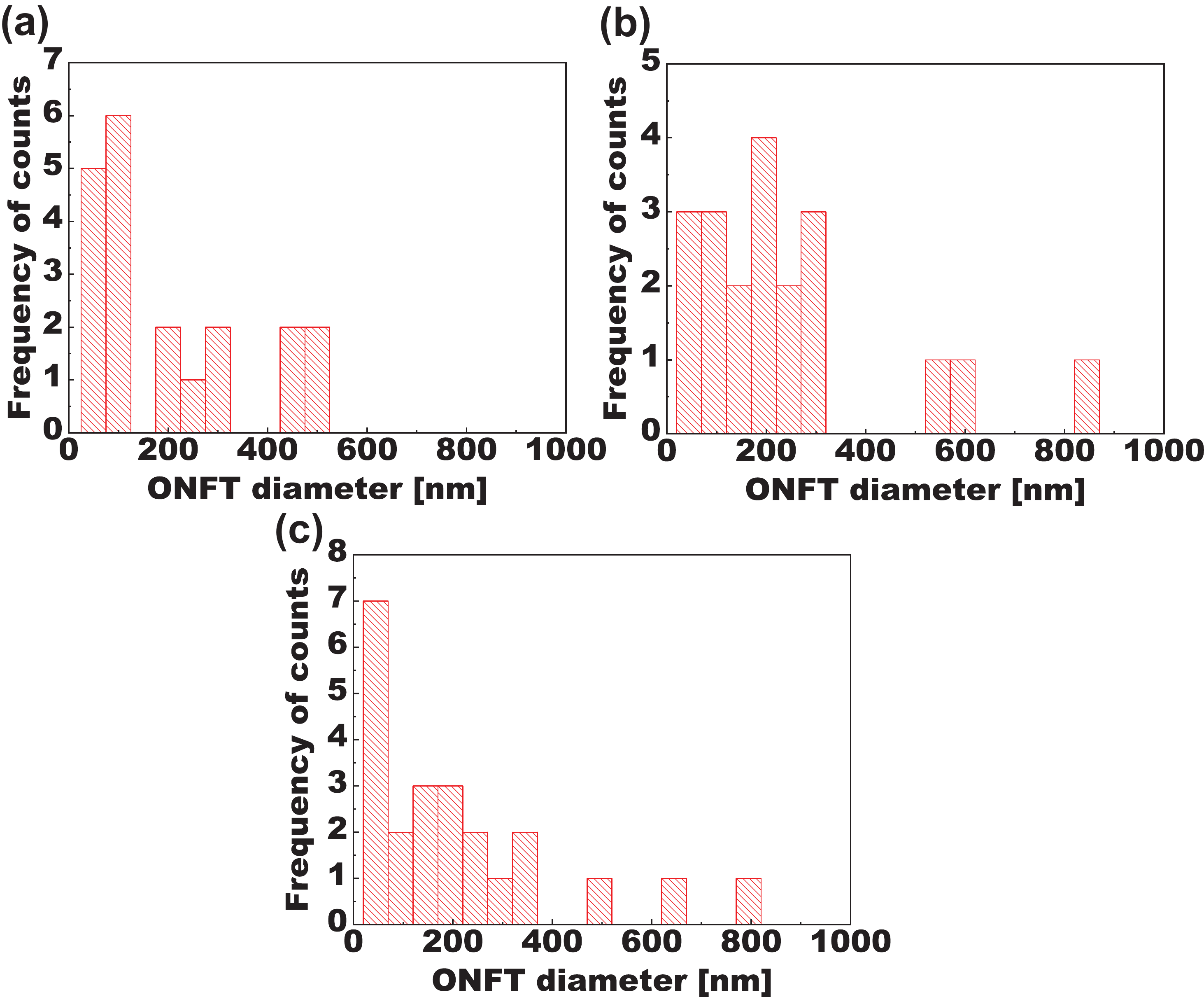}
	\caption{Histogram profiles for measured diameters of optical nanofiber tips with different etching timings of (a) 90 and 60 minutes (b) 90 and 70 minutes, and (c) 90 and 80 minutes.}
	\label{fig4}
\end{figure}

\begin{figure}[!h]
	\centering
	\includegraphics[width=\linewidth]{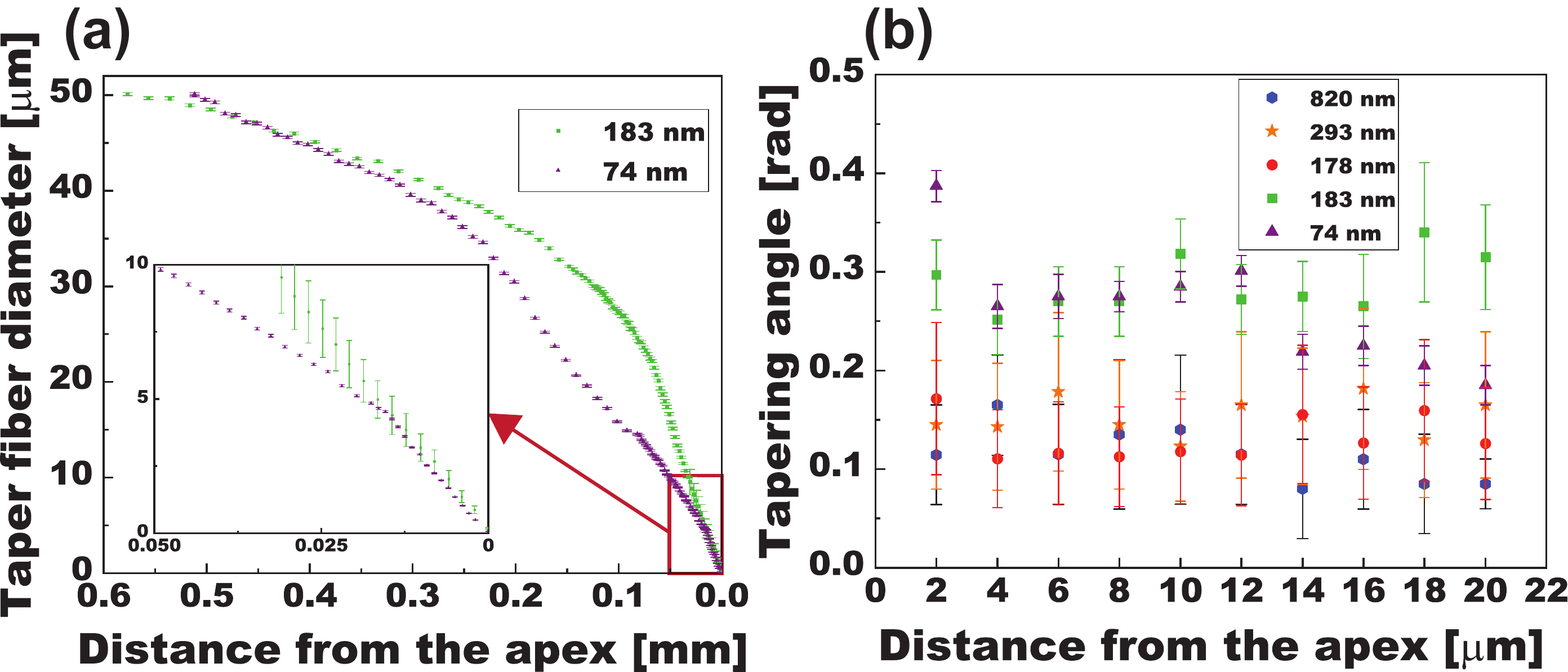}
	\caption{(a) Typical tapering profiles as a function of distance from the apex of ONFTs of tip diameters 183 $nm$ and 74 $nm$. Inset shows the tapering profile upto fiber diameter of 10 $\mu m$. (b) Tapering angle as a function of distance from the apex of ONFTs with diameters 820 $nm$, 293 $nm$, and 178 $nm$, 183 $nm$ and 74 $nm$.}
	\label{fig5}
\end{figure}

We measured ONFT diameters for several samples, and summaries for different etching times are plotted as histograms as shown in Figs. \ref{fig4} (a)-(c). Figure \ref{fig4} (a) represents the histogram for ONFT diameters for the etching timings 90 and 60 minutes for the first and second steps, respectively. The minimum/maximum measured ONFT diameter is 26/498 $nm$. Figure \ref{fig4} (b) represents the histogram for ONFT diameters for the etching timings 90 and 70 minutes for the first and second steps, respectively. The minimum/maximum measured ONFT diameter is 27/820 $nm$. Figure \ref{fig4} (c) represents the histogram for ONFT diameters for the etching timings 90 and 80 minutes for the first and second steps, respectively. The minimum/maximum ONFT diameter is 22/806 $nm$. Note that bin widths of histograms are 50 $nm$, similar to measurement errors. 

We also measured diameters of the ONFT at different locations in the tapering region along the fiber axis. Typical tapering profiles as a function of distance from the tip of the ONFT are shown in Fig. \ref{fig5} (a). Note that tip diameters are 183 $nm$ and 74 $nm$ and maximum diameter is 50.1 $\mu m$ for both. The fiber diameter of 50 $\mu m$ is achieved at a distance of $\sim$0.5 $mm$ from the tip. We analysed diameter measurements to investigate tapering angle. The tapering angle is calculated as the ratio of the difference in diameters at two positions along the fiber axis and distance between them. The summary of tapering angle as a function of the distance from the tip is shown in Fig. \ref{fig5} (b). The summary includes tapering angle profile for the three ONFTs shown in Figs. \ref{fig3} (c)-(e). and the two ONFTs mentioned in Fig. \ref{fig5} (a). The tapering angle varies from 0.08 $rad$ to 0.38 $rad$.
%\pagebreak
\section{Discussions}
\label{sec:Disc}

As seen in Fig. \ref{fig2} (b), the diffraction pattern is the first preview of the fabricated ONFT. The measured result agrees with the previously reported result in Ref. \cite{zhu2013dynamic}. This may indicate the smooth tapering and the micro/nano dimension of the ONFT. The measured optical transmission data from 36\% to 95\% shown in Fig. \ref{fig2} (d) indicates that ONFTs fabricated using this technique show good optical quality. Most samples exhibit optical transmission of $\ge$ 60\%, confirming that the chemical etching employed here is an efficient technique for fabricating optically good ONFTs. A summary of typical optical transmissions with corresponding ONFT diameter can be found in Table. \ref{tab1}. Diameter and optical transmission of previously reported results are also shown for comparison. Note that the SMF patch cable used to guide the laser light into the ONFT acts as a single mode filter. Therefore the measured transmission is for the single mode. Hence higher transmission is observed. As one can see in Table. \ref{tab1}, some samples show poor transmission. This may be due to the multimode excitation in tapering region. Large variation of optical transmissions might be due to the tapering profile and surface roughness. Since the transmission is measured, it can be accounted in further experiments for applications. We usually use optically good samples for further experiments.
\begin{table}[htb]
	\centering
	\caption{Summary of typical ONFT diameters and corresponding optical transmission}\label{tab1}%
	\begin{tabular}{|c|c|c|}
		\hline
		S. No. & Diameter of the ONFT ($nm$) & Optical transmission (\%)\\
		\hline
		1&217 ($\pm $10)&36 ($\pm $1)\\
		2 & 204 ($\pm $10)& 46 ($\pm $2)\\
		3&830 ($\pm $70)&65 ($\pm $2.5)\\
        4&274 ($\pm $20)&90 ($\pm $1.5)\\
		5&22 ($\pm $7)&91 ($\pm $3)\\
		6&820 ($\pm $50)&47 ($\pm$1.5)\\
		7&582 ($\pm $40)&75 ($\pm $1.5)\\
		8&127 ($\pm $15)&95 ($\pm $1)\\
	    9&293 ($\pm $70)&60 ($\pm $2)\\
	   10&178 ($\pm $50)&81 ($\pm $1)\\
	   \hline
	   11&820 (Ref. \cite{lee2019fabrication})&90\\
	   12&3000 (Ref.\cite{ward2006heat})&93\\
		\hline
	\end{tabular}
\end{table}

After the first step, as seen in Figs. \ref{fig3} (a) and (b), the measured diameters are 45 $\mu m$ and 47 $\mu m$, respectively. The measured average etching rate for the first step is 0.8 $\mu m/min$. As seen in Figs. \ref{fig3} (c)-(f), the measured diameters of ONFTs vary from 820 $nm$ to 22 $nm$. The measured average etching rate for the second step is 0.55 - 0.75 $\mu m/min$. The fluctuations in diameters may be due to the change in temperature and humidity as these levels directly influence the etching rate. We recorded temperature during the etching process and adjusted etching times acordingly. Although we followed 90 minutes as the etching time for the first step for temperature in the range of 25-28$^{\circ}$C, we reduced it to 60 minutes when the temperature was $\ge$ 28$^{\circ}$C at the beginning of the first step. The etching time for the second step was reduced to 30/40 minutes accordingly to achieve required fiber diameter. Also, the SMF size was reduced to only a few micrometers in humid condition and for temperature in the range of 23-25$^{\circ}$C. Note that the stability of environmental parameters is required to ensure consistent results. As these structures are so fragile, there is a chance that the tip may be broken while withdrawing from the HF. Additionally, a small jerk also cause breaking of the tip during experiment. At the end, the quality of the tip is asserted by measuring the transmission and diffraction pattern for further experiments.   
 
 The fabrication of ONFTs with a wide range of diameters is confirmed using the present technique. Note that ONFTs with minimum diameters of a few tens of nanometers are experimentally realized. The trend of the histograms for three particular etching times as shown in Figs. \ref{fig4} (a)-(c) for the measured ONFT diameters shows that the ONFT diameter does not vary as expected. We fabricate ONFTs with diameters from tens to hundreds of nanometers for all the three timings. The ONFT diameter does not decrease as the etching time is increased. Though the ONFT diameters cannot be correlated to etching timings, the present method is consistent to fabricate ONFTs of sizes less than 1 $\mu m$. At this point of time, our lab is not equipped to investigate and precisely control other parameters which affect the fabrication. In future we are planning to control these parameters and to quantitatively study the effects. To have precise control over the fabrication process, we are exploring heat and pull technique \cite{stiebeiner2010design,nayak2018nanofiber,hoffman2014ultrahigh} as another alternate approach. Consistent tip diameter and transmission is promised in this technique. 
 
 The tapering profile shown in Fig. \ref{fig5} (a), indicates that the fiber diameter reaches 50 $\mu m$ at a distance of 0.5 $mm$ from the tip of the ONFT. We dipped 2 $mm$ of the fiber in the HF for fabrication. This implies that the actual fiber diameter of 125 $\mu m$ may be achieved within 2 $mm$. Additionally, to improve the optical transmission with short tapering length, we investigate different dipping lengths of the SMF in the HF acid. We dipped the SMF into the HF acid up to 1.5 $mm$ and 1 $mm$. As the length dipped in the HF gets reduced, shorter tapered ONFTs with good optical quality can be realized. Such short-tapered ONFTs can be used in sensing and quantum optics \cite{lee2011planar}. Tapering profiles for the two samples follow almost the same trend. It indicates gradual tapering of the fiber, leading to high optical transmission. The tapering angle shown in Fig. \ref{fig5} (b), varies from 0.08 $rad$ to 0.38 $rad$. It indicates that the tapering is not smooth compared to heat and pull technique \cite{stiebeiner2010design}, in which tapering angle varies in $mrad$. Various samples do not follow the same trend. Although it does not follow the trend as that of heat and pull technique, chemically fabricated ONFTs are good enough to perform experiments. Regarding the two-step etching, if 40\% HF is used for the entire etching process, the tapering will be very sharp and there will be difficulty in controlling the size of the ONFT at nanoscales. If 24\% HF is used for the entire process, the etching process will take long time and the environmental conditions will change considerably. Hence, a two-step etching is preferred over a single step etching due to the control in etching rate.    

Regarding the surface smoothness of the ONFT, as seen in Figs. \ref{fig3}, roughness and crystalline structures are readily observed on the surface. Also, there is a non-uniform etching throughout the fiber, especially in low temperatures. Therefore, uncertainty in the ONFT diameter is inevitable in this technique. To realize high-quality ONFTs, one can extend the fabrication of the ONFT using the heat and pull technique \cite{stiebeiner2010design,nayak2018nanofiber,hoffman2014ultrahigh}. In this technique, accurate control over the ONFT diameter can be achieved. However, there is a trade-off between the tip diameter and tapering length. Therefore, depending on the application, one can adopt either chemical etching or heat and pull technique. 

 Although experimental ambiguities exist, we experimentally demonstrated the fabrication of ONFTs with optical transmissions from 36\% to 95\% and tip diameters ranging from 22 $nm$ to 820 $nm$. Note that always the tapered ONFT is in submicron scale, which is suitable for applications. ONFTs sizes less than 1 $\mu m$ is good enough for applications according to simulations \cite{resmichanneling,chonan2014efficient,resmi2023efficient,das2023efficient}. ONFTs with smaller diameters is efficient for sensing applications \cite{shashank,berneschi2020optical,ascorbe2016high,zaca2018etched,chyad2015fabrication,paiva2018optical,pathak2017fabrication}. Recently, our group reported channeling of fluorescence photons from quantum dots into guided modes of the ONFT which was fabricated using this technique \cite{resmichanneling}. The channeling was confirmed by measuring the photon-counting and emission spectrum of fluorescence photons. To calculate the channeling efficiency, optical transmission is significant. Also measurement of tip diameter is essential to determine whether the experiments are being carried out in the single mode regime or multi mode regime. Hence ONFTs will find more applications in quantum optics. 

\section{Conclusion}
\label{sec:Con}

In summary, we demonstrated the development of an optical nanofiber tip (ONFT) using a two-step chemical etching technique. We characterized fabricated ONFTs by measuring optical transmission and surface morphology. The observed optical transmissions are more than $30\%$ and tip diameters are less than $500$ $nm$. Temperature, humidity and other environmental parameters have to be maintained constant for consistent results. Consequently, tapering profile and surface roughness can minimize variation in optical transmissions. The measured results readily reveal the merit of the employed technique as easy and cost-effective. Due to the strong confinement of the electromagnetic field at the tip of ONFTs, these structures lay versatile platforms with potential applications in sensing, photonics, and quantum optics.
% \disclosures 
\subsection*{\textbf{Disclosures}}
The authors declare no conflict of interests.

\subsection* {\textbf{Code, Data, and Materials Availability}} 
Data may be obtained from authors upon reasonable request.

\subsection* {\textbf{Acknowledgments}}
This work was carried out in the wet chemistry lab under UGC-NRC in the School of Physics, University of Hyderabad. The authors thank Shashank Suman and Bratati Das for their helpful discussions. RM acknowledges the University Grants Commission (UGC) for the financial support (Ref. No.:1412/CSIR-UGC NET June 2019). RRY acknowledges the Science and Engineering Research Board (SERB) for the Core Research Grant (CRG) (File No. CRG/2021/009185) and the Institute of Eminence (IoE) grant at the University of Hyderabad, Ministry of Education (MoE) (File No. RC2-21-019).

%%%%% References %%%%%

\bibliography{report}   % bibliography data in report.bib
\bibliographystyle{spiejour}   % makes bibtex use spiejour.bst

%%%%% Biographies of authors %%%%%

\vspace{2ex}\noindent\textbf{Ramachandrarao Yalla} is an assistant professor at the University of Hyderabad, India. He received his MSc in Physics from the University of Hyderabad, India, in 2008 and his Ph.D. in Physics from the University of Electro-Communications, Japan, in 2012. He is the author of more than 15 journal papers, including Physical Review Letters and Applied Physics Letters. His current research interests include quantum optics, nanophotonics, and optoelectronic systems.

\vspace{2ex}\noindent\textbf{Resmi M} is a PhD scholar at the University of Hyderabad, India. She received her MSc degree in Physics from University of Madras, India in 2016. She is the author of 5 articles including journal paper and conference paper. 

\vspace{2ex}\noindent\textbf{Elaganuru Bashaiah} is a PhD scholar at the University of Hyderabad, India. He received his MSc degree in Physics from Vikrama Simhapuri University, India in 2016. He is the author of 5 articles including journal and conference papers.
%\vspace{1ex}
%\noindent Biographies and photographs of the other authors are not available.

%\listoffigures
%\listoftables

\end{spacing}
\end{document}